\documentclass[12pt]{article}
\usepackage{epsfig,graphics,graphicx,graphpap,color}
\usepackage{setspace}
\newcommand\bmat{\left( \begin{array}{cc}}
\newcommand\emat{\end{array}\right)}

\newcommand{\mat}[2]{\left(\begin{array}{#1} #2 \end{array}\right)}
\def\msbar{\ifmmode{\overline{\rm MS}} \else{$\overline{\rm MS}$} \fi}
\def\drbar{\ifmmode{\overline{\rm DR}} \else{$\overline{\rm DR}$} \fi}
\def\beq      {\begin{equation}}
\def\eeq      {\end{equation}}
\def\ti              {\tilde}

\def\e               {\epsilon}

\def\x               {\chi}

\newcommand{\neu}[1]   {{\ti \x^0_{#1}}}

\def\drbarp{\ifmmode{\overline{\rm DR'}} \else{$\overline{\rm DR}'$} \fi}

\begin{document}
\pagestyle{empty} \vspace*{-1cm}
\begin{flushright}
  HEPHY-PUB 831/06 \\
\end{flushright}
\vspace*{2cm}
\begin{center}
{\Large\bf\boldmath
   Two-loop SUSY QCD corrections to the neutralino masses in the MSSM}
   \\[5mm]
\vspace{10mm}
$\mbox{R. Sch\"ofbeck  and H. Eberl}$\\[5mm]
\vspace{6mm} $$ \mbox{{\it Institut f\"ur
Hochenergiephysik der \"Osterreichischen Akademie der
Wissenschaften,}}$$\vspace{-0.9cm} $$\mbox{{\it A--1050 Vienna,
Austria}}$$
\end{center}

\vspace{20mm}
\begin{abstract}
We have calculated the two-loop strong interaction corrections to the neutralino pole masses in the \drbarp scheme in the
Minimal Supersymmetric Standard Model (MSSM). We have performed a detailed numerical analysis for a particular point in the parameter space
and found corrections of a few tenths of a percent. We agree with previously derived analytic formulae for two-loop corrections to fermion masses.
\end{abstract}
\vfill
\newpage
\pagestyle{plain} \setcounter{page}{2}
\section{Introduction}
If supersymmetry (SUSY) as the most attractive extension of the
Standard Model is realized at low energies, the next generation of
high energy physics experiments at Tevatron, LHC and a future
$e^+e^-$ linear collider will discover supersymmetric particles.
Particularly at a linear collider, it will be possible to perform
measurements with high precision \cite{tesla, lincol, Weiglein:2004hn} which allows us
to test the underlying SUSY model.
According to work done by the LHC/LC study group \cite{Weiglein:2004hn}, the experimental accuracies
for the determination of the neutralino masses of $m_{\tilde\chi_1^0}$ and $m_{\tilde\chi_2^0}$ are 50~MeV and 80~MeV,
respectively, at the SPS1a' benchmark point \cite{spa}. To match this
accuracy it is indispensable to include higher order radiative
corrections.

In the framework of the Standard Model many calculations of quark pole masses were performed to two-loop \cite{Gray:1990yh}-\cite{Jegerlehner:2003py}
and three-loop order \cite{Chetyrkin:1999qi,Melnikov:2000qh}. Electroweak vector bosons were also studied in \cite{Chang:1981qq}-\cite{Jegerlehner:2002em} in this
respect. Within the MSSM important results on quark self-energies were obtained in \cite{Bednyakov:2002sf}-\cite{Bednyakov:2005kt}. Finally, in \cite{Martin:2005ch,Yamada:2006vn}
the gluino pole mass to two-loop order was calculated.

General formulae for SUSY QCD corrections to fermion pole masses in the MSSM were derived in \cite{Martin:2005ch}. In this paper, we
calculate the two-loop strong interaction corrections to the neutralinos within the MSSM in a completely independent way. We make a detailed numerical
analysis for the benchmark point SPS1a'. We agree with the analytic formulae given in \cite{Martin:2005ch}.

\section{Outline of the calculation}
The poles of the full renormalized propagator of a system of
mixing Majorana fermions in the mass eigenbasis
\beq
       G^{-1}_{\neu i\neu j}(k^2) = -i \Gamma^{(2)}_{ij}(k^2) = -i\mat{cc}{-M_D+\hat\Sigma_m^{LL}(k^2)&\sigma\cdot k (1+\hat\Sigma_{  k}^R(k^2)) \\
                     \bar\sigma\cdot k(1+\hat\Sigma_{k}^L(k^2)) & - M_D^\dagger+\hat\Sigma_m^{RR}(k^2)}_{ij}\label{defSE}
\eeq
in terms of the renormalized self-energies $\hat\Sigma_m^{LL}$, $\hat\Sigma_m^{RR}$, $\hat\Sigma_{  k}^L$ and $\hat\Sigma_{  k}^R$
 satisfy the complex pole mass equation
\beq
0=\det(s-(M_D-\hat\Sigma_m^{LL})\cdot(1+\hat\Sigma_{  k}^L)^{-1}\cdot(M_D^\dagger-\hat\Sigma_m^{RR})\cdot(1+\hat\Sigma_{ k}^{R})^{-1}),
\eeq
$M_D=\textrm{diag}(m_{\tilde\chi_1^0},m_{\tilde\chi_2^0},m_{\tilde\chi_3^0},m_{\tilde\chi_4^0})$. The solution of this equation up
to the considered order $s_{pole,i} = \hat m_{ii}^2- i \Gamma_i \hat m_{ii} =(M_{Dii}+\delta m_i)^2$ determines the
complex mass shift $\delta m_i$ and thus the width $\Gamma_i$ and the real part of the pole $\hat m_i$.
In a perturbative expansion to  $\mathcal{O}(\alpha\alpha_S)$ for the case of neutralinos in the MSSM the shifts in terms
 of self-energies are written as
\begin{eqnarray}
 \delta m_i & = & \delta m^{(1)}_i+\delta m^{(2)}_i\nonumber\\
            & = & - \frac{1}{2}(\hat\Sigma_m^{(1)LL}{}_i{}^i
                          + \hat\Sigma_m^{(1)RR}{}_i{}^i
                          + \hat\Sigma_k^{(1)L}{}_i{}^i  M_{Dii}
                          +\hat\Sigma_k^{(1)R}{}_i{}^i
                          M_{Dii})\nonumber\\
              && - \frac{1}{2}(\hat\Sigma_m^{(2)LL}{}_i{}^i
                          + \hat\Sigma_m^{(2)RR}{}_i{}^i
                          + \hat\Sigma_k^{(2)L}{}_i{}^i  M_{Dii}
                          +\hat\Sigma_k^{(2)R}{}_i{}^i
                          M_{Dii})\label{EQOLTLcorr}.
\end{eqnarray}
where there is no sum on the index $i$. Note that there are no one-loop squared or
one-loop differentiated contributions in eq. (\ref{EQOLTLcorr}) at the two-loop level because $\alpha_S$ is absent at one-loop level.
\\
\begin{figure}[h]
\begin{center}
\begin{picture}(125,50)(0,0)
    \put(-19,-8){\mbox{\resizebox{!}{5cm}{\includegraphics{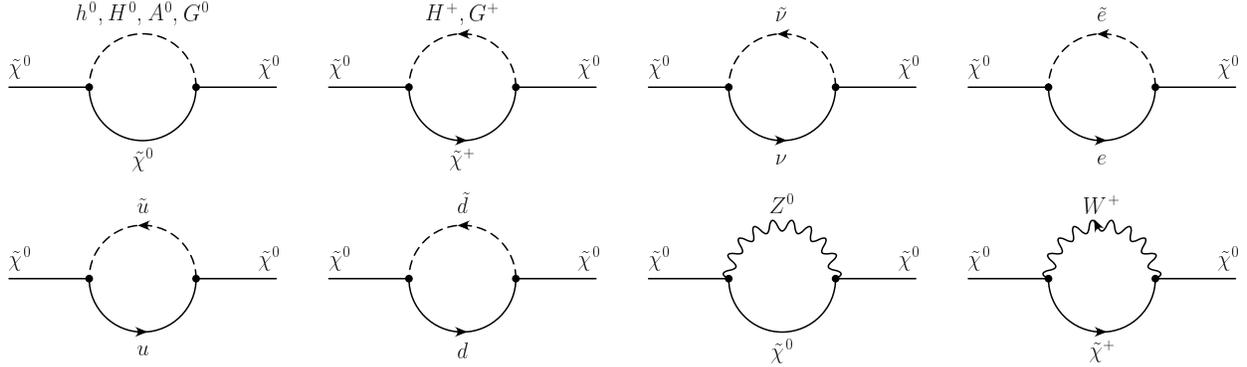}}}}
\end{picture}
\end{center}
\caption{\it Neutralino one-loop self-energy diagrams}\label{NE1LoopDiags}
\end{figure}

\begin{figure}[h!]
\begin{center}
\begin{picture}(125,25)(0,0)
    \put(-19,-8){\mbox{\resizebox{!}{2.95cm}{\includegraphics{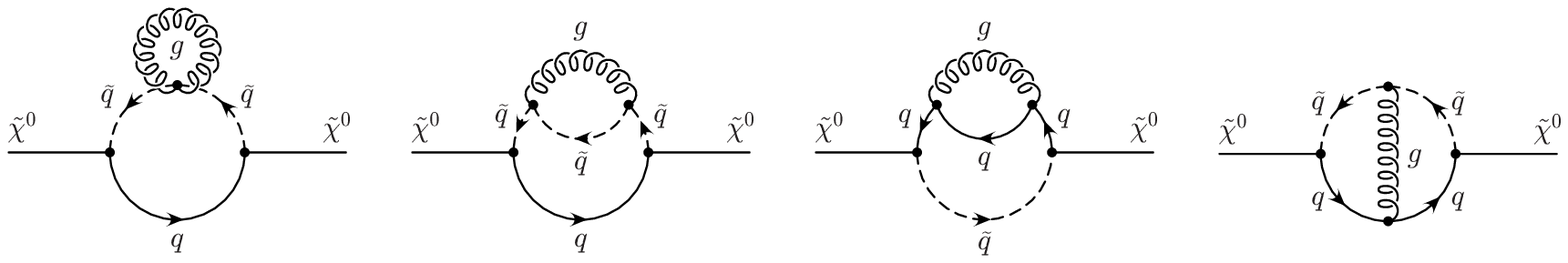}}}}
\end{picture}
\end{center}
\caption{\it Neutralino two-loop selfenergy diagrams with inner gluon line}\label{NE2LGluon}
\end{figure}

\begin{figure}[h]
\begin{center}
\begin{picture}(125,25)(0,0)
    \put(-4,-9){\mbox{\resizebox{!}{2.80cm}{\includegraphics{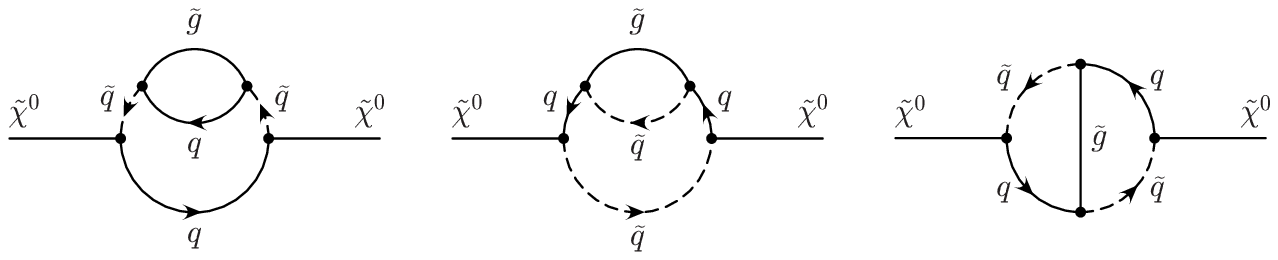}}}}
\end{picture}
\end{center}
\caption{\it  Neutralino two-loop selfenergy diagrams with inner gluino line}\label{NE2LGluino}
\end{figure}

\begin{figure}[h]
\begin{center}
\begin{picture}(125,30)(0,0)
   \put(20,-10){\mbox{\resizebox{!}{3.3cm}{\includegraphics{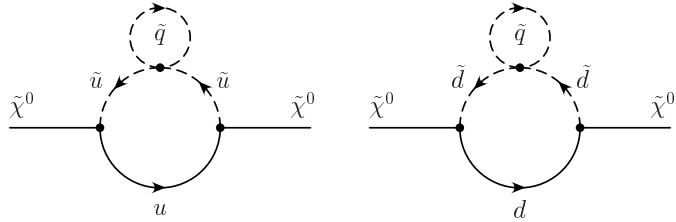}}}}
\end{picture}
\end{center}
\caption{\it  Neutralino two-loop selfenergy diagrams with three inner squark lines}\label{NE2LSquark}
\end{figure}

\begin{figure}[h!]
\begin{center}
\begin{picture}(125,25)(0,0)
    \put(-19,-8){\mbox{\resizebox{!}{2.95cm}{\includegraphics{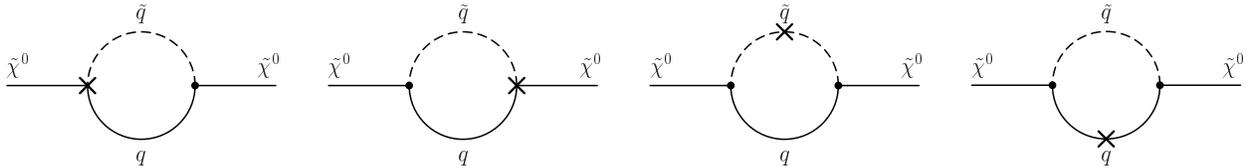}}}}
\end{picture}
\end{center}
\caption{\it  Neutralino two-loop self-energy diagrams with counter term insertions}\label{NECT}
\end{figure}

In Fig.~\ref{NE1LoopDiags} we show the one-loop diagrams. Note that diagrams with charge conjugated inner particles are not shown explicitly.
We checked our analytic one-loop calculation against previous work  \cite{Oller:2003ge, Fritzsche:2004ek} in the on-shell scheme and found agreement.

The nine different types of two-loop diagrams stem from SUSY QCD operator
insertions into the quark/squark neutralino one-loop self-energy
and  can be grouped into those containing inner gluon
(Fig.~\ref{NE2LGluon}), gluino (Fig.~\ref{NE2LGluino}) and squark
lines (Fig.~\ref{NE2LSquark}). The latter only contains the part of the 4-squark interaction
proportional to $\alpha_S$.
Inserting $(u,d)$ for every quark line in these
diagrams gives a total number of 36 explicit two-loop diagrams to be calculated.
Diagrams with one-loop counter term insertions (Fig.~\ref{NECT}) involve $\mathcal O (\alpha_S)$ mass counter terms for quarks and squarks as well as
coupling constant counter terms stemming from the Yukawa part of the neutralino-quark-squark coupling and counter terms to the squark mixing matrix,
see e.g. \cite{Bednyakov:2002sf}.

The diagrams and the amplitudes were generated in {\sc FeynArts 3.2} \cite{feynarts} and
simplified analytically using {\sc FeynCalc} 4.0.2 \cite{feyncalc}. A small IR regulator mass parametrized artificial IR divergencies, and it was checked
analytically that they cancel  each other in the final result. According to the \drbarp-scheme the UV-divergencies are regulated dimensionally
by $d=4-\e$, and the unphysical scalar mass parameter $m_\e$ for the evanescent fields is absorbed according to \cite{Martin:2001vx}.
The resulting integrals have a rather complex tensor structure which was reduced to a set of basis integrals
using {\sc Tarcer} \cite{tarcer}. The resulting expressions are, however, too lengthy to be given here.
The numerical analysis was then performed by implementing  {\sc Tsil} \cite{tsil} in a {\sc Fortran} program which was auto-generated with the help
of {\sc Mathematica}.
Apart from the check of gauge independence, we used the usual 't Hooft Feynman $R_{\xi=1}$ gauge for the gluon field.
\section{Numerical results}

\begin{figure}[h!]
\begin{center}
\resizebox{!}{5cm}{\includegraphics{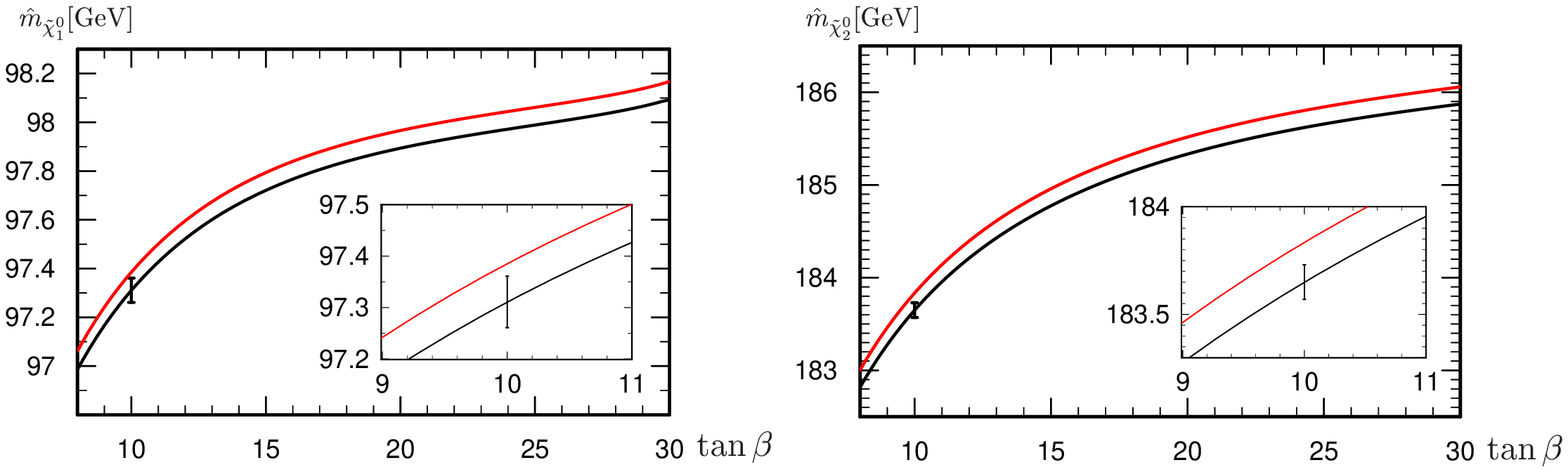}}
 \caption{\it Neutralino pole masses  $\hat m_{\neu 1}$ and $\hat m_{\neu 2}$ as a funtion of $\tan\beta$ with experimental accuracy at SPS1a'.
 The red and the black line
 are the one- and two-loop results, respectively.}\label{tanb}
 \end{center}
\end{figure}
\begin{figure}[h!]
\begin{center}
\resizebox{!}{10cm}{\includegraphics{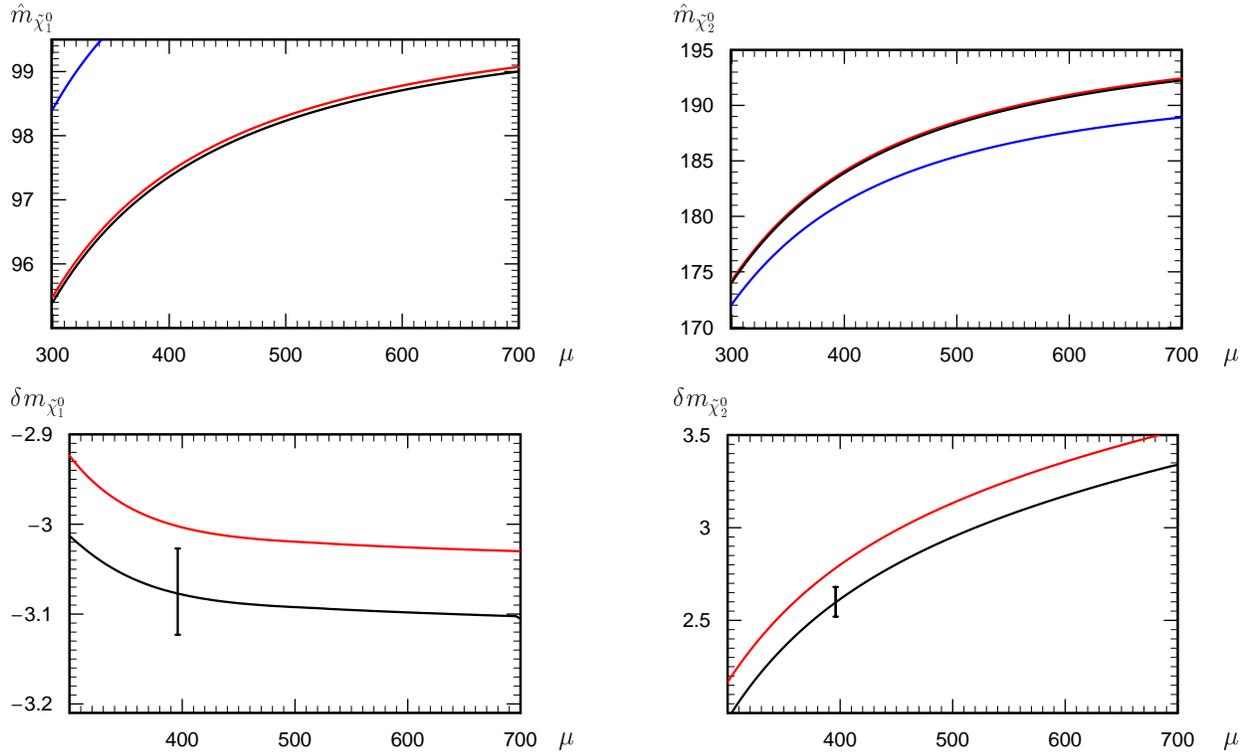}}
 \caption{\it Neutralino pole masses as a function of  $\mu$ for $\neu 1$ and $\neu 2$ and the respective absolute corrections, all numbers in GeV.
 The red and the black line
 are the one- and two-loop results, respectively. Blue denotes the tree-level \drbar masses.}\label{mu}
 \end{center}
\end{figure}

\begin{figure}[h!]
\begin{center}
\resizebox{!}{6cm}{\includegraphics{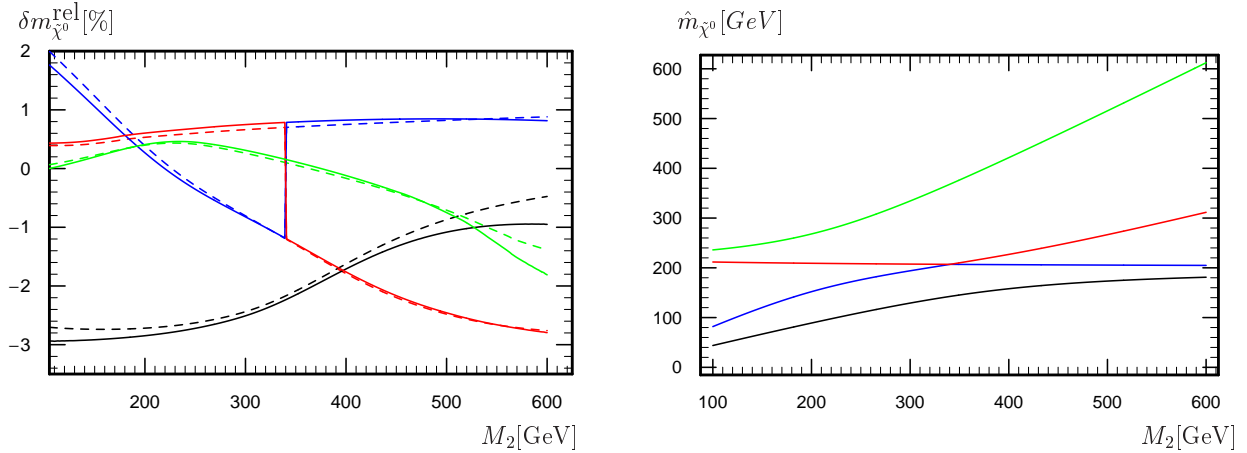}}
 \caption{\it Left: relative corrections assuming gauge unification. Dashed and solid lines are one- and two-loop corrections, respectively. Right: Tree-level \drbar masses in this scenario.
 The black line is for  $\neu 1$,  blue  for $\neu 2$, red  for $\neu 3$ and green for $\neu 4$. Notice the level crossing between
 $\neu 2$ and $\neu 3$ at approx. 350~GeV.}\label{gaugeuni}
 \end{center}
\end{figure}

\begin{figure}[h!]
\begin{center}
\resizebox{!}{5.8cm}{\includegraphics{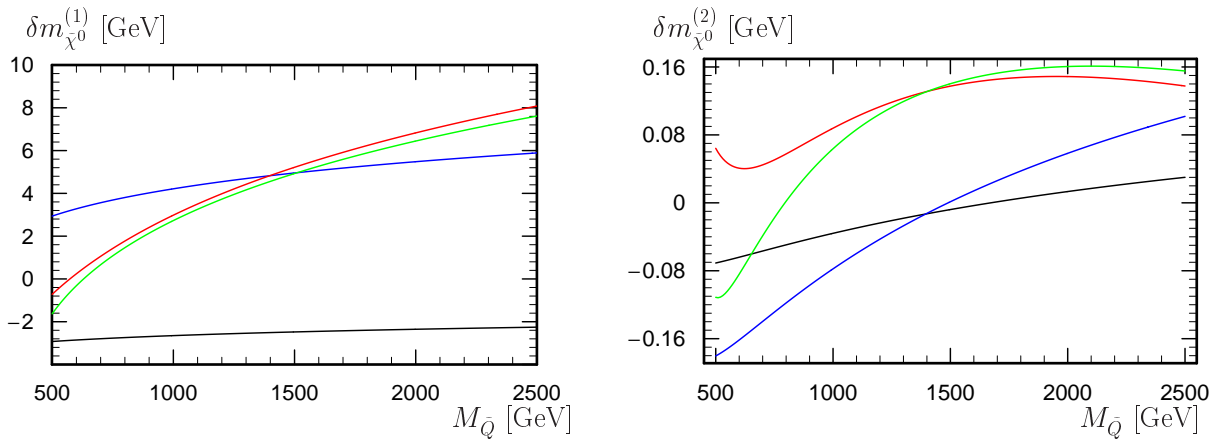}}
 \caption{{\it Absolute neutralino mass shifts as a function of $M_{\ti Q}$, see text.
  All other parameters from SPS1a'.
 Left: absolute one-loop corrections.  Right: absolute two-loop corrections. The black lines are for $\neu 1$,
 blue  for $\neu 2$, red  for $\neu 3$ and green for $\neu 4$. }}\label{MSQ3}
 \end{center}
\end{figure}

\begin{figure}[t]
\begin{center}
 \includegraphics{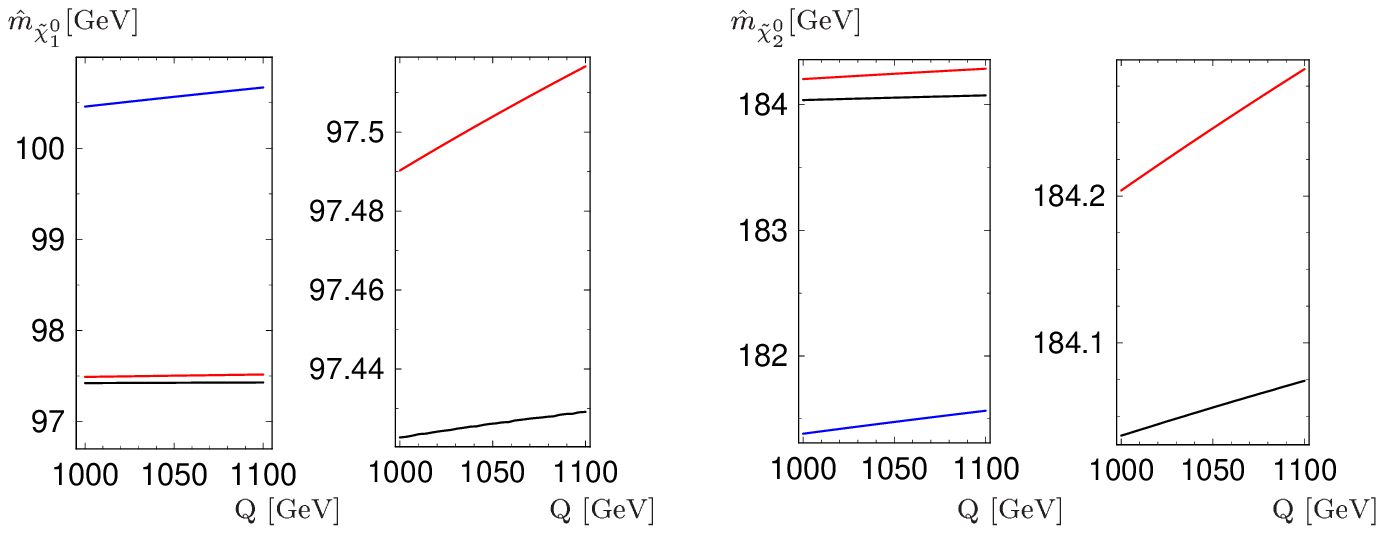}
 \caption{\it \it Scale dependence of $\hat m_{\neu 1}$ and $\hat m_{\neu 2}$ at SPS1a'. The blue line is the tree-level \drbar
 input value, red and black are the pole masses at one- and two-loop level, respectively.}\label{rendep}
 \end{center}
\end{figure}

For the numerical analysis, we took the benchmark point SPS1a' \cite{spa}. The SUSY parameters at $Q_0=1$~TeV are $\tan\beta = 10$,
$M_1 = 103.209$~GeV, $M_2 = 193.295$~GeV, $M_3 = 572.328$~GeV, $\mu = 401.62$~GeV, $A_t = -532.38$~GeV, $A_b = -938.91$~GeV,
$M_{\ti Q_3} = 470.91$~GeV, $M_{\ti U_3} = 385.32$~GeV and $M_{\ti D_3} = 501.37$~GeV, for further details see again \cite{spa}.

Fig.~\ref{tanb} shows the neutralino pole masses at SPS1a' except for $\tan\beta$ which is varied between 3 and 30.
At the SPS1a' value $\tan\beta=10$ the corrections are larger than the experimental accuracy.

The two-loop corrections to the masses of $\neu 3$ and $\neu 4$ are of the order of $0.5$~GeV lying far beyond the expected eperimental accuracy of
$2-3$~GeV \cite{Weiglein:2004hn}.

In Fig.~\ref{mu} we vary $\mu$ from 300~GeV to 700~GeV. Since the tree-level neutralino masses strongly depend on this parameter we show
the absolute one- and two-loop corrections in seperate plots.

A rather different scenario is shown in Fig.~\ref{gaugeuni} assuming gauge unification. The plot is over $M_2$, $\mu$ is set to $200$~GeV and
all other values are taken from SPS1a'. Taking over experimental accuracy from SPS1a' which is 0.05\%  for $m_{\neu 1}$ we again
find important corrections in this scenario (one tick on the vertical axis is 0.2\%).

In Fig.~\ref{MSQ3} we show the one- and two-loop neutralino mass shifts as a function of the third generation soft SUSY breaking masses $M_{\ti Q}=M_{\ti Q_3}=M_{\ti U_3}=M_{\ti D_3}$.
Again, all other parameters are taken from SPS1a'.

\section{Checks}
We have checked analytically the gauge-$\xi$  independence of the sum of those diagrams containing inner gluon lines  in the 't Hooft gauge.
The diagrams containing gluon lines provide $\e$-scalar mass contributions which were treated  separately and shown to drop out by
 redefining the soft bilinear squark breaking parameters \cite{Martin:2001vx}.

The scaling behaviour given by the $\textrm{log}[Q]$ and $\textrm{log}^2[Q]$ structure of our results was
checked against the renormalization group prediction \cite{Martin:1993zk} analytically and numerically.
For the latter, Fig.~\ref{rendep} illustrates the improvement of the scale dependence for the pole masses of $\neu 1$ and $\neu 2$.
The two plots with only a red and a black line are
zooms of the respective plots to their left in order to illustrate graphically the improvement when going from one- to two-loop level.
For this plot we used the RGE subroutines of {\sc Spheno} \cite{Porod:2003um} which were extended to include also running standard model parameters.
The remaining scale dependence of the neutralino pole masses indicate that further non-SUSY QCD corrections like Yukawa-interactions
will give relevant contributions.

The analytic RGE check is equivalent to the UV finiteness in the on-shell scheme. The trace of the square of the tree-level \drbar mass matrix,
\begin{equation}\label{eqdefTdrbar}
       T^\drbar = \textrm{tr}(Y^\dagger Y)^\drbar = \textrm{tr}(M_D^\dagger M_D)^\drbar = (M_1^2+M_2^2+2(\mu^2+M_Z^2))^\drbar = \sum_{i=1}^4m^2_{\ti\chi^0_i}{}^\drbar\,,
\end{equation}
is a sum of squared mass parameters and involves the whole set of two-loop self-energies when
renormalized. In fact, it was shown that the sum of the squares of the pole masses is scale independent,
\begin{equation}\label{eqdefT}
0 = \frac{\textrm{d} }{\textrm{d} t}\left.\textrm{tr}\left[(M_D(t)+\delta m(X(t)))^\dagger(M_D(t)+\delta m(X(t)))\right]\right|_{\alpha\alpha_S}\,,
\end{equation}
order by order in pertubation theory. $X(t)$ abbreviates the set of \drbar input parameters $X^\drbar$ run to the scale Q.
First, differentiating and expanding $T(t)=\textrm{tr}(Y(t)^\dagger Y(t))$ around $t = \log[Q/Q_0]=0$ with $T(0)=T^\drbar$
 up to order $\alpha\alpha_S$ and $t\,\alpha\alpha_S$ one has to evaluate
\begin{equation}\label{eqrunT}
       \frac{\textrm{d} T(t)}{\textrm{d}t}\simeq \frac{\partial T}{\partial X}\left.\left(\frac{1}{16\pi^2}\beta_X^{(1)}+\frac{1}{(16\pi^2)^2}\left(t\frac{\partial\beta_X^{(1)}}{\partial X'}\beta^{(1)}_{X'}+\beta_X^{(2)}\right)\right)\right|_{X^\drbar}
\end{equation}
using the respective one- and two-loop $\beta$-functions $\beta^{(1,2)}_X$  given in \cite{Martin:1993zk}
where sums over the indices of $X$ and $X'$ are understood implicitly. At the benchmark point SPS1a' one has $Q_0=1~\textrm{TeV}$.
The absence of strong interactions at the one-loop level is reflected by the fact
that the $\alpha_S$ factor in $t\,\alpha\alpha_S$ only stems from $\beta^{(1)}_{X'}$ in eq.~(\ref{eqrunT}). Second, the rhs of eq.~(\ref{eqrunT}) has to be
compensated by the $Q$-dependence of the mass shifts $\delta m_i$ in eq.~(\ref{eqdefT}) evaluated with running parameters $X(t)$.
Note that in eq.~(\ref{eqdefT}) the one-loop shift $\delta m_i^{(1)}$ has to be differentiated with respect to all $\alpha_S$-running parameters.
Again, one-loop squared terms cannot contribute due to the absence of $\alpha_S$ at the one-loop level.

As a last check, an independent C-program was written to evaluate the formulae given in \cite{Martin:2005ch} which agrees numerically with our auto-generated code
up to the seventh relevant digit.

\section{Conclusions and Outlook}
Combining existing software packages \cite{feynarts,feyncalc,tarcer,tsil} we have calculated the two-loop corrections
in $\mathcal{O}(\alpha\alpha_S)$ to the neutralino pole masses in the MSSM in the \drbarp scheme.
We have shown that the results are important for matching theory and experiment at future colliders. Many checks have
 been succesful and the renormalization scale dependence is improved.
However, it seems to be necessary to include further corrections, e.g. the leading two-loop Yukawa corrections $\mathcal{O}(\alpha h_t^2,\alpha h_b^2)$.

\clearpage
{\bf Acknowledgements}\\
The authors would like to thank W.~Majerotto, K.~Kova\v{r}\'{\i}k and
C.~Weber for discussion and many useful comments throughout the last year.
They especially thank W. Majerotto for his help in finalizing this work.
The authors acknowledge support from EU under the MRTN-CT-2006-035505
network programme. This work is supported by the "Fonds zur F\"orderung
der wissenschaftlichen Forschung" of Austria, project No. P18959-N16.

\end{document}